\documentclass[sigconf]{acmart}
\usepackage{booktabs} 
\usepackage{multirow}
\usepackage{color}
\usepackage{subfigure} 
\usepackage{array}
\usepackage{breqn}
\usepackage{{inputenc}}
\usepackage{balance}
\usepackage{multirow,tabularx}
\usepackage{longtable}
\usepackage{booktabs,longtable}
\usepackage{hhline}
\usepackage[flushleft]{threeparttable}
\usepackage{algorithm}
\usepackage{algorithmic}
\usepackage{epsfig}
\usepackage{graphicx}
\usepackage{epstopdf}
\usepackage{thmtools}
\usepackage{enumitem}
\usepackage{verbatim}
\usepackage{amsfonts}
\usepackage{hyperref}
\hypersetup{colorlinks=false,linkcolor=blue,urlcolor=blue,citecolor=red}
\newcommand\norm[1]{\left\lVert#1\right\rVert}

\newcommand{\lapl}{\mathcal{L}}
\newcommand{\Trans}[1]{{#1}^{\top}}

\newcommand{\Mat}[1]{\mathbf{#1}}

\newcommand{\Space}[1]{\mathbb{#1}}
\newcommand{\Set}[1]{\mathcal{#1}}

\newcommand{\ParaTitle}[1]{\noindent\textbf{#1}}

\newcommand{\ie}{\emph{i.e., }}
\newcommand{\eg}{\emph{e.g., }}

\newcommand{\wrt}{\emph{w.r.t. }}
\newcommand{\cf}{\emph{cf. }}
\newcommand{\aka}{\emph{aka. }}

\AtBeginDocument{%
  \providecommand\BibTeX{{%
    \normalfont B\kern-0.5em{\scshape i\kern-0.25em b}\kern-0.8em\TeX}}}

\copyrightyear{2020}
\acmYear{2020}
\setcopyright{acmcopyright}
\acmConference[SIGIR '20]{Proceedings of the 43rd International ACM SIGIR Conference on Research and Development in Information Retrieval}{July 25--30, 2020}{Virtual Event, China}
\acmBooktitle{Proceedings of the 43rd International ACM SIGIR Conference on Research and Development in Information Retrieval (SIGIR '20), July 25--30, 2020, Virtual Event, China}
\acmPrice{15.00}
\acmDOI{10.1145/3397271.3401137}
\acmISBN{978-1-4503-8016-4/20/07}

\begin{document}
\fancyhead{}

\title{Disentangled Graph Collaborative Filtering}
\author{Xiang Wang}
\affiliation{%
	\institution{National University of Singapore}
}
\email{xiangwang@u.nus.edu}

\author{Hongye Jin}
\affiliation{%
	\institution{Peking University}
}
\email{mooler0410@gmail.com}

\author{An Zhang}
\affiliation{%
	\institution{National University of Singapore}
}
\email{anzhang@u.nus.edu}

\author{Xiangnan He}
\affiliation{%
	\institution{University of Science and Technology of China}
}
\email{xiangnanhe@gmail.com}

\author{Tong Xu}
\affiliation{%
	\institution{University of Science and Technology of China}
}
\email{tongxu@ustc.edu.cn}

\author{Tat-Seng Chua}
\affiliation{%
	\institution{National University of Singapore}
}
\email{dcscts@nus.edu.sg}


\begin{abstract}
	Learning informative representations of users and items from the interaction data is of crucial importance to collaborative filtering (CF).
	Present embedding functions exploit user-item relationships to enrich the representations, evolving from a single user-item instance to the holistic interaction graph.
	Nevertheless, they largely model the relationships in a uniform manner, while neglecting the diversity of user intents on adopting the items, which could be to pass time, for interest, or shopping for others like families.
	Such uniform approach to model user interests easily results in suboptimal representations, failing to model diverse relationships and disentangle user intents in representations.
	

	In this work, we pay special attention to user-item relationships at the finer granularity of user intents.
	We hence devise a new model, \emph{Disentangled Graph Collaborative Filtering} (DGCF), to disentangle these factors and yield disentangled representations.
	Specifically, by modeling a distribution over intents for each user-item interaction, we iteratively refine the intent-aware interaction graphs and representations.
	Meanwhile, we encourage independence of different intents.
	This leads to disentangled representations, effectively distilling information pertinent to each intent.
	We conduct extensive experiments on three benchmark datasets, and DGCF achieves significant improvements over several state-of-the-art models like NGCF~\cite{NGCF}, DisenGCN~\cite{DisenGCN}, and MacridVAE~\cite{MacridVAE}.
	Further analyses offer insights into the advantages of DGCF on the disentanglement of user intents and interpretability of representations.
	Our codes are available in \url{https://github.com/xiangwang1223/disentangled_graph_collaborative_filtering}.
	
\end{abstract}

\ccsdesc[500]{Information systems~Recommender systems}

\keywords{Collaborative Filtering, Graph Neural Networks, Disentangled Representation Learning, Explainable Recommendation}

\maketitle

\section{Introduction}

Personalized recommendation has become increasingly prevalent in real-world applications, to help users in discovering items of interest.
Hence, the ability to accurately capture user preference is the core.
As an effective solution, collaborative filtering (CF), which focuses on historical user-item interactions (\eg purchases, clicks), presumes that behaviorally similar users are likely to have similar preference on items.
Extensive studies on CF-based recommenders have been conducted and achieved great success.

\begin{figure}[th]
    \centering
	\includegraphics[width=0.47\textwidth]{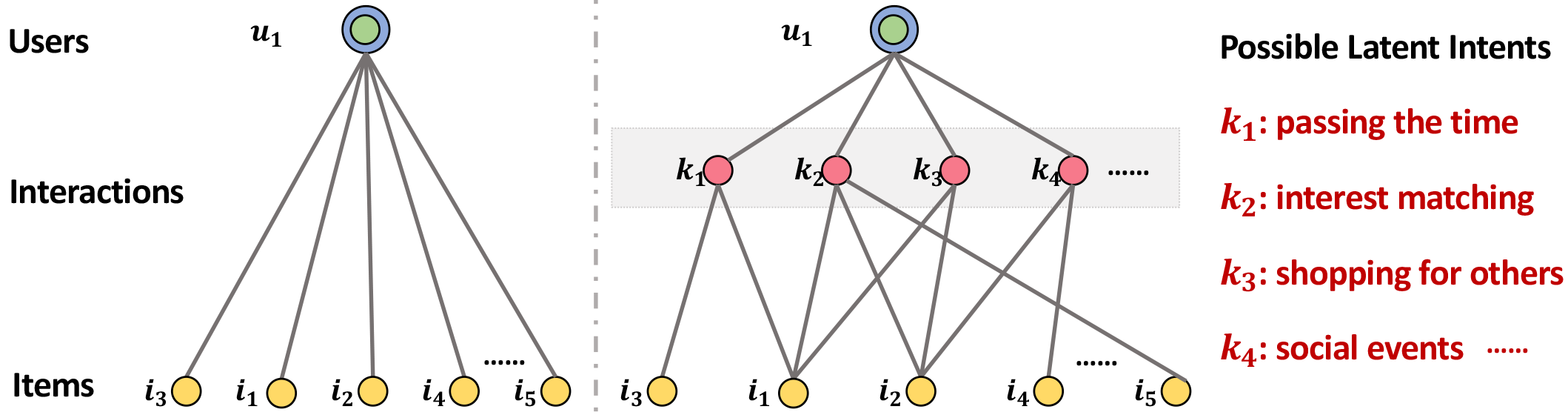}
	\vspace{-5pt}
	\caption{An illustration of diverse user-item relationships at the granularity of latent intents.}
	\label{fig:intro}
	\vspace{-5pt}
\end{figure}

Learning informative representations of users and items is of crucial importance to improving CF.
To this end, the potentials of deepening user-item relationships become more apparent.
Early models like matrix factorization (MF)~\cite{BPRMF} forgo user-item relationships in the embedding function by individually projecting each user/item ID into a vectorized representation (\aka embedding).
Some follow-on studies~\cite{SVDFeatures,NAIS,Mult-VAE,FISM} introduce personal history as the pre-existing feature of a user, and integrate embeddings of historical items to enrich her representation.
More recent works~\cite{NGCF,GC-MC,lightgcn} further organize all historical interactions as a bipartite user-item graph to integrate the multi-hop neighbors into the representations and achieved the state-of-the-art performance.
We attribute such remarkable improvements to the modeling of user-item relationships, evolving from using only a single ID, to personal history, and then holistic interaction graph.

Despite effectiveness, we argue that prior manner of modeling user-item relationships is insufficient to discover disentangled user intents.
The key reason is that existing embedding functions fail to differentiate user intents on different items --- they either treat a user-item interaction as an isolated data instance~\cite{BPRMF,NCF} or uniformly organize it as an edge in the interaction graph~\cite{NGCF,GC-MC,lightgcn} (as shown in the left of Figure~\ref{fig:intro}) to train the neural networks.
An underlying fact is omitted that: a user generally has multiple intents to adopt certain items; moreover, different intents could motivate different user behaviors~\cite{DisenGCN,MacridVAE,GNNExplainer,DBLP:conf/icde/ChenYCYNL19}.
Taking the right of Figure~\ref{fig:intro} as an example, user $u$ watches movie $i_{1}$ to pass time, while cares less about whether $i_{1}$'s attributes (\eg director) match with her interests; on the other hand, $u$'s interaction with $i_{2}$ may be mainly driven by her special interests on its director.
Leaving this fact untouched, previous modeling of user-item relationships is coarse-grained, which has several limitations:
1) Without considering the actual user intents could easily lead to suboptimal representations;
2) As noisy behaviors (\eg random clicks) commonly exist in a user's interaction history, confounding her intents makes the representations less robust to noisy interactions;
and 3) User intents will be obscure and highly entangled in representations, which results in poor interpretability.


Having realized the vital role of user-item relationships and the limitations of prior embedding functions, we focus on exploring the relationships at a more granular level of user intents, to disentangle these factors in the representations.
Intuitively, there are multiple intents affecting a user's behavior, such as to pass time, interest matching, or shopping for others like families.
We need to learn a distribution over user intents for each user behavior, summarizing the confidence of each intent being the reason why a user adopts an item.
Jointly analyzing such distributions of all historical interactions, we can obtain a set of intent-aware interaction graphs, which further distill the signals of user intents.
However, this is not trivial due to the following challenges:
\begin{itemize}[leftmargin=*]
    \item How to explicitly present signals pertinent to each intent in a representation is unclear and remains unexplored;
    \item The quality of disentanglement is influenced by the independence among intents, which requires a tailored modeling.
\end{itemize}

In this work, we develop a new model, \emph{Disentangled Graph Collaborative Filtering} (DGCF), to disentangle representations of users and items at the granularity of user intents.
In particular, we first slice each user/item embedding into chunks, coupling each chunk with a latent intent.
We then apply an \textbf{graph disentangling module} equipped with neighbor routing~\cite{CapsuleNetwork,DisenGCN} and embedding propagation~\cite{GCN,GraphSAGE,NGCF,KGAT} mechanisms.
More formally, neighbor routing exploits a node-neighbor affinity to refine the intent-aware graphs, highlighting the importance of influential relationships between users and items.
In turn, embedding propagation on such graphs updates a node's intent-aware embedding.
By iteratively performing such disentangling operations, we establish a set of intent-aware graphs and chunked representations.
Simultaneously, an \textbf{independence modeling module} is introduced to encourage independence of different intents. Specifically, a statistic measure, distance correlation~\cite{szekely2007measuring,szekely2009brownian}, is employed on intent-aware representations.
As the end of these steps, we obtain the disentangled representations, as well as explanatory graphs for intents.
Empirically, DGCF is able to achieve better performance than the state-of-the-art methods such as NGCF~\cite{NGCF}, MacridVAE~\cite{MacridVAE}, and DisenGCN~\cite{DisenGCN} on three benchmark datasets.

We further make in-depth analyses on DGCF's disentangled representations \wrt disentanglement and interpretability.
To be more specific, we find that the discovered intents serve as \emph{vitamin} to representations --- that is, even in small quantities could achieve comparable performance, while the deficiency of any intent would hinder the results severely. Moreover, we use side information (\ie user reviews) to help interpret what information are being captured in the intent-aware graphs, trying to understand the semantics of intents.

In a nutshell, this work makes the following main contributions:
\begin{itemize}[leftmargin=*]
    \item We emphasize the importance of diverse user-item relationships in collaborative filtering, and modeling of such relationships could lead to better representations and interpretability.
    
    \item We propose a new model DGCF, which considers user-item relationships at the finer granularity of user intents and generates disentangled representations.
    
    \item We conduct extensive experiments on three benchmark datasets, to demonstrate the advantages of our DGCF on the effectiveness of recommendation, disentanglement of latent user intents, and interpretability of representations.
\end{itemize}

\section{Preliminary and Related Work}
We first introduce the representation learning of CF, emphasizing the limitations of existing works \wrt the modeling of user-item relationships.
In what follows, we present the task formulation of disentangling graph representations for CF.

\subsection{Learning for Collaborative Filtering}
Discovering user preference on items, based solely on user behavior data, lies in the core of CF modeling.
Typically, the behavior data involves a set of users $\Set{U}=\{u\}$, items $\Set{I}=\{i\}$, and their interactions $\Set{O}^{+}=\{y_{ui}\}$, where $y_{ui}=1$ is an observed interaction indicating that user $u$ adopted item $i$ before, otherwise $0$.
Hence, the prime task of CF is to predict how likely user $u$ would adopt item $i$, or more formally, the likelihood of their interaction $\hat{y}_{ui}$.

\subsubsection{\textbf{Learning Paradigm of CF}}
There exists a semantic gap between user and item IDs --- no overlapping between such superficial features --- hindering the interaction modeling.
Towards closing this gap, extensive studies have been conducted to learn informative representations of users and items.
Here we summarize the representation learning as follows:
\begin{gather}
    \Mat{e}_{u}=f(u),\quad\Mat{e}_{i}=f(i),
\end{gather}
where $u$ and $i$ separately denote the user and item IDs;
$\Mat{e}_{u}\in\Space{R}^{d}$ and $\Mat{e}_{i}\in\Space{R}^{d}$ are the vectorized representations (\aka embeddings) of user $u$ and item $i$, respectively; $f(\cdot)$ is the embedding function and $d$ is the embedding size.
Such embeddings are expected to memorize underlying characteristics of items and users.

Thereafter, interaction modeling is performed to reconstruct the historical interactions.
Inner product is a widely-used interaction function~\cite{BPRMF,NGCF,MF,HOP-rec}, which is employed on user and item representations to perform the prediction, as:
\begin{gather}\label{equ:inner-product}
    \hat{y}_{ui}=\Trans{\Mat{e}}_{u}\Mat{e}_{i},
\end{gather}
which casts the predictive task as the similarity estimation between $u$ and $i$ in the same latent space.
Here we focus on the representation learning of CF, thereby using inner product as the predictive model and leaving the exploration of interaction modeling in future work.

\subsubsection{\textbf{Representation Learning of CF}}
Existing works leverage user-item relationships to enrich CF representations, evolving from single user-item pairs, personal history to the holistic interaction graph.
Early, MF~\cite{BPRMF,MF,DBLP:conf/icml/SalakhutdinovMH07} projects each user/item ID into an embedding vector.
Many recommenders, such as NCF~\cite{NCF}, CMN~\cite{CMN}, and LRML~\cite{LRML} resort to this paradigm.
However, such paradigm treats every user-item pair as an isolated data instance, without considering their relationships in the embedding function.
Towards that, later studies like SVD++~\cite{SVDFeatures}, FISM~\cite{FISM}, NAIS~\cite{NAIS}, and ACF~\cite{ACF} view personal history of a user as her features, and integrate embeddings of historical items via the average~\cite{SVDFeatures,FISM} or attention network~\cite{NAIS,ACF} as user embeddings, showing promise in representation learning.
Another similar line applies autoencoders on interaction histories to estimate the generative process of user behaviors, such as Mult-VAE~\cite{Mult-VAE}, AutoRec~\cite{AutoRec}, and CDAE~\cite{CDAE}.
One step further, the holistic interaction graph is used to smooth or enrich representations.
Prior efforts like HOP-Rec~\cite{HOP-rec} and GRMF~\cite{GRMF} apply idea of unsupervised representation learning --- that is, connected nodes have similar representations --- to smooth user and item representations.
More recently, inspired by the great success of graph neural networks (GNNs)~\cite{GCN,GAT,GraphSAGE,SGC}, some works, such as GC-MC~\cite{GC-MC}, NGCF~\cite{NGCF}, PinSage~\cite{PinSage}, and LightGCN~\cite{lightgcn}, further reorganize personal histories in a graph, and distill useful information from multi-hop neighbors to refine the embeddings.

In the view of interaction graph, we can revisit such user-item relationships as the connectivity between user and item nodes.
In particular, bipartite user-item graph is denoted as $\Set{G}=\{(u,i,y_{ui})\}$, where the edge between user $u$ and item $i$ nodes is the observed interaction $y_{ui}$.
When exploring $\Set{G}$, connectivity is derived from a path starting from a user (or item) node, which carries rich semantic representing user-item relationships.
Examining paths rooted at $u_{1}$, we have the first-order connectivity $u_{1}\rightarrow \{i_{1},i_{2}\}$ showing her interaction history, which intuitively profiles her interests;
moreover, the second-order connectivity $u_{1}\rightarrow i_{1}\rightarrow u_{2}$ indicates behavioral similarity between $u_{1}$ and $u_{2}$, as both adopted $i_{1}$ before;
furthermore, we exhibit collaborative signals via the third-order connectivity $u_{1}\rightarrow i_{1}\rightarrow u_{2}\rightarrow i_{3}$, which suggests that $u_{1}$ is likely to consume $i_{3}$ since her similar user $u_{2}$ has adopted $i_{3}$.

As such, user-item relationships can be explicitly represented as the connectivity: single ID (\ie no connectivity)~\cite{BPRMF}, personal history (\ie the first-order connectivity)~\cite{SVDFeatures}, to holistic interaction graph (\ie higher-order connectivity)~\cite{NGCF,CF-UIcA,HOP-rec}.

\subsubsection{\textbf{Limitations}}
Despite of their success, we argue that such uniform modeling of user-item relationships is insufficient to reflect users' latent intents. This limiting the interpretability and understanding of CF representations.
To be more specific, present embedding functions largely employ a black-box neural network on the relationships (\eg interaction graph $\Set{G}$), and output representations.
They fail to differentiate user intents on different items by just presuming a uniform motivation behind behaviors.
However, it violates the fact that a user generally has multiple intents when purchasing items.
For example, user $u$ interacts with items $i_{1}$ and $i_{2}$ with intent to pass time and match personal taste, respectively.
However, such latent intents are less explored, and could easily lead to suboptimal representation ability.

Without modeling such user intents, existing works hardly offer interpretable embeddings, so as to understand semantics or what information are encoded in particular dimensions.
Specifically, the contributions of each interaction $(u,i)$ to all dimensions of $\Mat{e}_{u}$ are indistinguishable.
As a result, the latent intents behind each behavior are highly entangled in the embeddings, obscuring the mapping between intents and particular dimensions.

Study on disentangling representations for recommendation is less conducted until recent MacridVAE~\cite{MacridVAE}.
It employs $\beta$-VAE~\cite{BetaVAE} on interaction data and achieves disentangled representations of users.
Owing to the limitations of $\beta$-VAE, only distributions over historical items (\ie the first-order connectivity between users and items) are used to couple the discovered factors with user intents, while ignoring the complex user-item relationships (\ie higher-order connectivity reflecting collaborative signals).
We hence argue that the user-item relationships might be not fully explored in MacridVAE.

\subsection{Task Formulation}
We formulate our task that consists of two subtasks --- 1) exploring user-item relationships at a granular level of user intents and 2) generating disentangled CF representations.

\subsubsection{\textbf{Exploring User-Item Relationships.}}
Intuitively, one user behavior is influenced by multiple intents, such as passing time, matching particular interests, and shopping for others like family.
Taking movie recommendation as an example, user $u$ passed time with movie $i_{1}$, hence might care less about whether $i_{1}$'s director matches her interests well;
whereas, $u$ watched $i_{2}$ since its director is an important factor of $u$'s interest.
Clearly, different intents have varying contributions to motivate user behaviors. 

To model such finer-grained relationships between users and items, we aim to learn a distribution $A(u,i)$ over user intents for each behavior, as follows:
\begin{gather}\label{equ:intent-distribution-definition}
    \Mat{A}(u,i)=\Big(A_{1}(u,i),\cdots,A_{K}(u,i)\Big),
\end{gather}
where $A_{k}(u,i)$ reflects the confidence of the $k$-th intent being the reason why user $u$ adopts item $i$;
$K$ is the hyperparameter controlling the number of latent user intents.
Jointly examining the scores relevant to particular intent $k$, we can construct an intent-aware graph $\Set{G}_{k}$, which is defined as $\Set{G}_{k}=\{(u,i,A_{k}(u,i))\}$,
where each historical interaction $(u,i)$ represents one edge and is assigned with $A_{k}(u,i)$.
Moreover, a weighted adjacency matrix $\Mat{A}_{k}$ is built for $\Set{G}_{k}$.
As such, we establish a set of intent-aware graphs $\Set{G}=\{\Set{G}_{1},\cdots,\Set{G}_{K}\}$ to present diverse user-item relationships, instead of a uniform one adopted in prior works~\cite{NGCF,lightgcn,GC-MC}.

\begin{figure*}[th]
    \centering
	\includegraphics[width=0.82\textwidth]{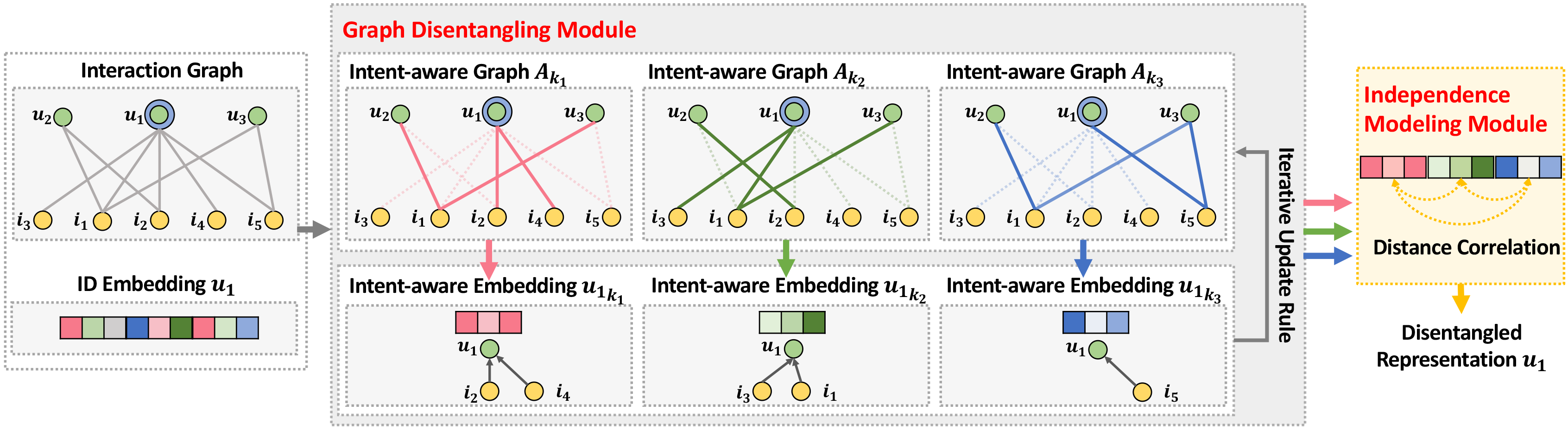}
	\vspace{-10pt}
	\caption{Illustration of the proposed disentangled graph collaborative filtering framework. Best viewed in color.}
	\label{fig:framework}
	\vspace{-10pt}
\end{figure*}

\subsubsection{\textbf{Generating Disentangled Representations.}}
We further target at exploiting the discovered intents to generate disentangled representations for users and items --- that is, extract information that are pertinent to individual intents as independent parts of representations.
More formally, we aim to devise an embedding function $f(\cdot)$, so as to output a disentangled representation $\Mat{e}_{u}$ for user $u$, which is composed of $K$ independent components:
\begin{gather}
    \Mat{e}_{u}=(\Mat{e}_{1u},\Mat{e}_{2u},\cdots,\Mat{e}_{Ku}),
\end{gather}
where $\Mat{e}_{ku}$ is the $k$-th latent intent influence for user $u$; for simplicity, we make these component the same dimension, $\Mat{e}_{ku}\in\Space{R}^{\frac{d}{K}}$.
It is worth highlighting that $e_{ku}$ should be independent of $\Mat{e}_{k'u}$ for $k'\neq k$,
so as to reduce semantic redundancy and encourage that signals are maximally compressed about individual intents.
Towards that, each chunked representation $\Mat{e}_{ku}$ is built upon the intent-aware graph $\Set{G}_{k}$ and synthesizes the relevant connectivities.
Analogously, we can establish the representation $\Mat{e}_{i}$ for item $i$.


\section{Methodology}
We now present disentangled graph collaborative filtering, termed DGCF, which is illustrated in Figure~\ref{fig:framework}.
It is composed of two key components to achieve disentanglement:
1) graph disentangling module, which first slices each user/item embedding into chunks, coupling each chunk with an intent, and then incorporates a new neighbor routing mechanism into graph neural network, so as to disentangle interaction graphs and refine intent-aware representations;
and 2) independence modeling module, which hires distance correlation as a regularizer to encourage independence of intents.
DGCF ultimately yields disentangled representations with intent-aware explanatory graphs.

\subsection{Graph Disentangling Module}
Studies on GNNs~\cite{GCN,GAT,GraphSAGE} have shown that applying embedding-propagation mechanism on graph structure can extract useful information from multi-hop neighbors and enrich representation of the ego node.
To be more specific, a node aggregates information from its neighbors and updates its representations.
Clearly, the connectivities among nodes provide an explicit channel to guide the information flow.
We hence develop a GNN model, termed \emph{graph disentangling layer}, 
which incorporates a new neighbor routing mechanism into the embedding propagation, so as to update weights of these graphs.
This allows us to differentiate varying importance scores of each user-item connection to refine the interaction graphs, and in turn propagate signals to the intent-aware chunks.

\subsubsection{\textbf{Intent-Aware Embedding Initialization}}
Distinct from mainstream CF models~\cite{BPRMF,NGCF,SVDFeatures,NCF} that parameterize user/item ID as a holistic representation only, we additionally separate the ID embeddings into $K$ chunks, associating each chunk with a latent intent.
More formally, such user embedding is initialized as:
\begin{gather}\label{equ:initial-chunked}
    \Mat{u}=(\Mat{u}_{1},\Mat{u}_{2},\cdots,\Mat{u}_{K}),
\end{gather}
where $\Mat{u}\in\Space{R}^{d}$ is ID embedding to capture intrinsic characteristics of $u$;
$\Mat{u}_{k}\in\Space{R}^{\frac{d}{K}}$ is $u$'s chunked representation of the $k$-th intent.
Analogously, $\Mat{i}=(\Mat{i}_{1},\Mat{i}_{2},\cdots,\Mat{i}_{K})$ is established for item $i$.
Hereafter, we separately adopt random initialization to initialize each chunk representation, to ensure the difference among intents in the beginning of training.
It is worth highlighting that, we set the same embedding size (say $d=64$) with the mainstream CF baselines, instead of doubling model parameters (\cf Section~\ref{sec:model-size}).

\subsubsection{\textbf{Intent-Aware Graph Initialization}}
We argue that prior works are insufficient to profile rich user intents behind behaviors, since they only utilize one user-item interaction graph~\cite{NGCF} or homogeneous rating graphs~\cite{GC-MC} to exhibit user-item relationships.
Hence, we define a set of score matrices $\{\Mat{S}_{k}|\forall k\in\{1,\cdots,K\}\}$ for $K$ latent intents.
Focusing on an intent-aware matrix $\Mat{S}_{k}$, each entry $S_{k}(u,i)$ denotes the interaction between user $u$ and item $i$.
Furthermore, for each interaction, we can construct a score vector $\Mat{S}(u,i)=(S_{1}(u,i),\cdots,S_{K}(u,i))\in\Space{R}^{K}$ over $K$ latent intents.
We uniformly initialize each score vectors as follows:
\begin{gather}\label{equ:intial-intent-distribution}
    \Mat{S}(u,i)=(1,\cdots,1),
\end{gather}
which presumes the equal contributions of intents at the start of modeling.
Hence, such score matrix $\Mat{S}_{k}$ can be seen as the adjacency matrix of intent-aware graph.

\subsubsection{\textbf{Graph Disentangling Layer}}\label{sec:graph-disentangling}
Each intent $k$ now includes a set of chunked representations, $\{\Mat{u}_{k},\Mat{i}_{k}|u\in\Set{U},i\in\Set{I}\}$, which specialize its feature space, as well as a specific interaction graph represented by $\Mat{S}_{k}$.
Within individual intent channels, we aim to distill useful information from high-order connectivity between users and items, going beyond ID embeddings.
Towards this end, we devise a new graph disentangling layer, which is equipped with the neighbor routing and embedding propagation mechanisms, with the target of differentiating adaptive roles of each user-item connection when propagating information along it.
We define such layer $g(\cdot)$ as follows:
\begin{gather}\label{equ:first-order-propagation}
    \Mat{e}^{(1)}_{ku}=g(\Mat{u}_{k},\{\Mat{i}_{k}|i\in\Set{N}_{u}\}),
\end{gather}
where $\Mat{e}^{(1)}_{ku}$ is to collect information that are pertinent to intent $k$ from $u$'s neighbors;
$\Set{N}_{u}$ is the first-hop neighbors of $u$ (\ie the historical items adopted by $u$);
and the super-index $(1)$ denotes the first-order neighbors.

\begin{figure}[t]
    \centering
	\includegraphics[width=0.47\textwidth]{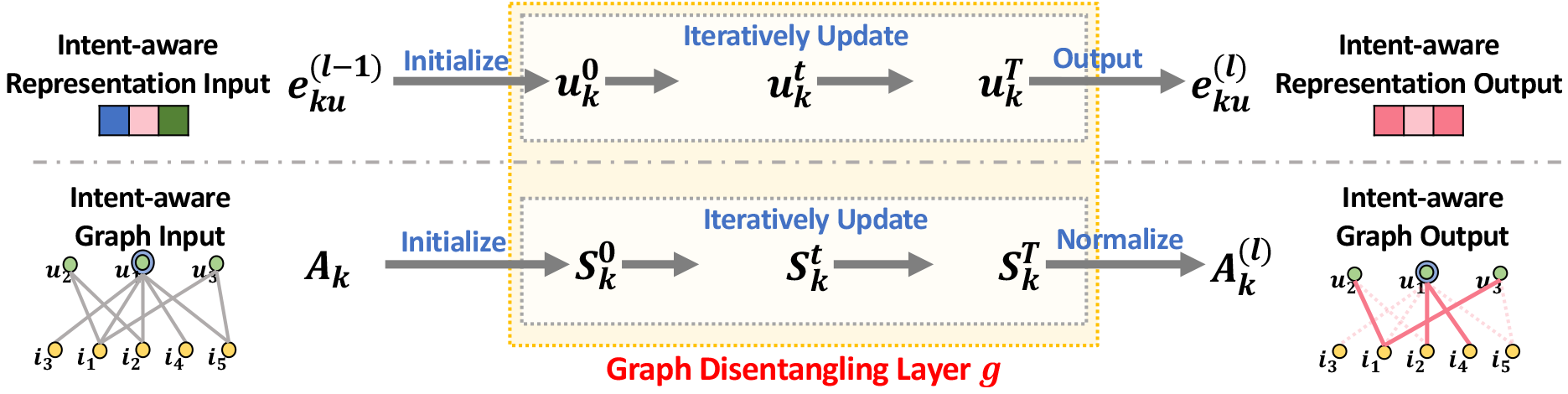}
	\vspace{-10pt}
	\caption{Illustration of iterative update rule.}
	\label{fig:iterative-rule}
	\vspace{-15pt}
\end{figure}

\vspace{5pt}
\ParaTitle{Iterative Update Rule.}
Thereafter, as Figure~\ref{fig:iterative-rule} shows, the neighbor routing mechanism is adopted: first, we employ the embedding propagation mechanism to update the intent-aware embeddings, based on the intent-aware graphs; then, we in turn utilize the updated embeddings to refine the graphs and output the distributions over intents.
In particular, we set $T$ iterations to achieve such iterative update.
In each iteration $t$, $\Mat{S}^{t}_{k}$ and $\Mat{u}^{t}_{k}$ separately memorize the updated values of adjacency matrix and embeddings, where $t\in\{1,2,\cdots,T\}$ and $T$ is the terminal iteration.
It starts by initializing $\Mat{S}^{0}_{k}=\Mat{S}_{k}$ and $\Mat{u}^{0}_{k}=\Mat{u}_{k}$ via Equations~\eqref{equ:intial-intent-distribution} and~\eqref{equ:initial-chunked}.


\vspace{5pt}
\ParaTitle{Cross-Intent Embedding Propagation.}
At iteration $t$, for the target interaction $(u,i)$, we have the score vector, say $\{\Mat{S}_{k}(u,i)|\forall k\in\{1,\cdots,K\}\}$.
To obtain its distribution over all intents, we then normalize these coefficients via the softmax function:
\begin{gather}
    \tilde{S}^{t}_{k}(u,i)=\frac{\exp{S^{t}_{k}(u,i)}}{\sum_{k'=1}^{K}\exp{S^{t}_{k'}(u,i)}},
\end{gather}
which is capable of illustrating which intents should get more attention to explain each user behavior $(u,i)$.
As a result, we can get the normalized adjacency matrix $\tilde{\Mat{S}}^{t}_{k}$ for each intent $k$.
We then perform embedding propagation~\cite{GCN,GAT,GraphSAGE} over individual graphs, such that the information, which are influential to the user intent $k$, are encoded into the representations.
More formally, the weighted sum aggregator is defined as:
\begin{gather}\label{equ:embedding-propagation}
    \Mat{u}^{t}_{k}=\sum_{i\in\Set{N}_{u}}\lapl^{t}_{k}(u,i)\cdot\Mat{i}^{0}_{k},
\end{gather}
where $\Mat{u}^{t}_{k}$ is $u$'s temporary representation to memorize signals refined from her neighbors $\Set{N}_{u}=\{i|(u,i)\in\Set{G}\}$, after $t$ iterations; $\Mat{i}^{0}_{k}$ is the input representation for historical item $i$; and 
and $\lapl^{t}_{k}(u,i)$ is the Laplacian matrix of $\tilde{\Mat{S}}^{t}_{k}$, formulated as:
\begin{gather}
    \lapl^{t}_{k}(u,i)=\frac{\tilde{S}^{t}_{k}(u,i)}{\sqrt{D^{t}_{k}(u)\cdot D^{t}_{k}(i)}},
\end{gather}
where $D^{t}_{k}(u)=\sum_{i'\in\Set{N}_{u}}\tilde{S}^{t}_{k}(u,i')$ and $D^{t}_{k}(i)=\sum_{u'\in\Set{N}_{i}}\tilde{S}^{t}_{k}(u',i)$ are the degrees of user $u$ and item $i$, respectively; $N_{u}$ and $N_{i}$ are the one-hop neighbors of $u$ and $i$, respectively.
Obviously, when iteration $t=1$, $D^{1}_{k}(u)$ and $D^{1}_{k}(i)$ separately degrade as $|\Set{N}_{u}|$ and $|\Set{N}_{i}|$, which is the fixed decay term widely adopted in prior studies~\cite{GCN,SGC}.
Such normalization can handle the varying neighbor numbers of nodes, making the training process more stable.

It is worth emphasizing that, we aggregate the initial chunked representations $\{\Mat{i}^{0}_{k}\}$ as the distilled information for user $u$.
This contains the signal from the first-order connectivities only, while excluding that from user $u$ herself and her higher-hop neighbors.
Moreover, inspired by recent SGC~\cite{SGC} and LightGCN~\cite{lightgcn}, we argue that nonlinear transformation adopted by prior works~\cite{NGCF,GC-MC} is burdensome for CF and its black-box nature hinders the disentanglement process, thereby omitting the transformation and using ID embeddings only.

\vspace{5pt}
\ParaTitle{Intent-Aware Graph Update.}
We iteratively adjust the edge strengths based on neighbors of a user (or an item) node.
Examining the subgraph structure rooted at user node $u$ with Equation~\eqref{equ:embedding-propagation}, $\Mat{u}^{t}_{k}$ can be seen as the centroid within the local pool $\Set{N}_{u}=\{(u,i)\}$, which contains items $u$ has interacted with before.
Intuitively, historical items driven by the same intent tend to have similar chunked representations, further encouraging their relationships to be stronger.
We hence iteratively update $S^{t}_{k}(u,i)$ --- more precisely, adjusting the strength between the centroid $u$ and its neighbor $i$, as follows:
\begin{gather}
    S^{t+1}_{k}(u,i)=S^{t}_{k}(u,i)+\Trans{\Mat{u}^{t}_{k}}\text{tanh}(\Mat{i}^{0}_{i}),
\end{gather}
where $\Trans{\Mat{u}^{t}_{k}}\text{tanh}(\Mat{i}^{0}_{k})$ considers the affinity between $\Mat{u}^{t}_{k}$ and $\Mat{i}^{0}_{k}$;
and $\text{tanh}$~\cite{Attention} is a nonlinear activation function to increase the representation ability of model.

After $T$ iterations, we ultimately obtain the output of one graph disentangling layer, which consists of disentangled representation \ie $\Mat{e}^{(1)}_{ku}=\Mat{u}^{T}_{k}$, as well as its intent-aware graph \ie $\Mat{A}^{(1)}_{k}=\tilde{\Mat{S}}^{T}_{k}$, $\forall k\in\{1,\cdots,K\}$.
When performing such propagation forward, our model aggregates information pertinent to each intent and generates an attention flow, which can be viewed as explanations behind the disentanglement. 

\subsubsection{\textbf{Layer Combination}}
Having used the first-hop neighbors, we further stack more graph disentangling layers to gather the influential signals from higher-order neighbors.
In particular, such connectivities carry rich semantics.
For instance, the second-order connectivity like $u_{1}\rightarrow i_{2}\rightarrow u_{3}$ suggests the intent similarity between $u_{1}$ and $u_{3}$ when consuming $i_{2}$;
meanwhile, the longer path $u_{1}\rightarrow i_{2}\rightarrow u_{2}\rightarrow i_{4}$ deepens their intents via the collaborative signal.
To capture user intents from such higher-order connectivity, we recursively formulate the representation after $l$ layers as:
\begin{gather}\label{equ:high-order-propagation}
    \Mat{e}^{(l)}_{ku}=g\Big(\Mat{e}^{(l-1)}_{ku},\{\Mat{e}^{(l-1)}_{ki}|i\in\Set{N}_{u}\}\Big),
\end{gather}
where $\Mat{e}^{(l-1)}_{ku}$ and $\Mat{e}^{(l-1)}_{ki}$ are the representations of user $u$ and item $i$ conditioned on the $k$-th factor, memorizing the information being propagated from their $(l-1)$-hop neighbors.
Moreover, each disentangled representation is also associated with its explanatory graphs to explicitly present the intents, \ie the weighted adjacency matrix $\Mat{A}^{(l)}_{k}$.
Such explanatory graphs are able to show reasonable evidences of what information construct the disentangled representations.

After $L$ layers, we sum the intent-aware representations at different layers up as the final representations, as follows:
\begin{gather}
    \Mat{e}_{ku}=\Mat{e}^{(0)}_{ku}+\cdots+\Mat{e}^{(L)}_{ku},\quad\quad\Mat{e}_{ki}=\Mat{e}^{(0)}_{ki}+\cdots+\Mat{e}^{(L)}_{ki}.
\end{gather}
By doing so, we not only disentangle the CF representations, but also have the explanations for each part of representations.
It is worthwhile to emphasize that the trainable parameters are only the embeddings at the $0$-th layer, \ie $\Mat{u}$ and $\Mat{i}$ for all users and items (\cf Equation~\eqref{equ:initial-chunked}).

\subsection{Independence Modeling Module}\label{sec:independence-modeling}
As suggested in~\cite{CapsuleNetwork,DisenGCN}, dynamic routing mechanism encourages the chunked representations conditioned on different intents to be different from each others.
However, the difference constraint enforced by dynamic routing is insufficient: there might be
redundancy among factor-aware representations.
For example, if one intent-aware representation $\Mat{u}_{k}$ can be inferred by the others $\{\Mat{u}_{k'}|k'\neq k\}$, the factor $k$ is highly likely to be redundant and could be confounding.

We hence introduce another module, which can hire statistical measures like mutual information~\cite{MINE} and distance correlation~\cite{szekely2007measuring,szekely2009brownian} as a regularizer, with the target of encouraging the factor-aware representations to be independent.
We here apply distance correlation, leaving the exploration of mutual information in future work.
In particular, distance correlation is able to characterize independence of any two paired vectors, from their both linear and nonlinear relationships; its coefficient is zero if and only if these vectors are independent~\cite{szekely2007measuring}.
We formulate this as:
\begin{gather}\label{equ:independence-loss}
    loss_{\text{ind}}=\sum^{K}_{k=1}\sum^{K}_{k'=k+1}dCor(\Mat{E}_{k},\Mat{E}_{k'}),
\end{gather}
where $\Mat{E}_{k}=[\Mat{e}_{u_{1}k},\cdots,\Mat{e}_{u_{N}k},\Mat{e}_{i_{1}k},\cdots,\Mat{e}_{i_{M}k}]\in\Space{R}^{(M+N)\times \frac{d}{K}}$ is the embedding look-up table with $N=|\Set{U}|$ and $M=|\Set{I}|$, which is built upon the intent-aware representations of all users and items; and, $dCor(\cdot)$ is the function of distance correlation defined as:
\begin{gather}
    dCor(\Mat{E}_{k},\Mat{E}_{k'})=\frac{dCov(\Mat{E}_{k},\Mat{E}_{k'})}{\sqrt{dVar(\Mat{E}_{k})\cdot dVar(\Mat{E}_{k'})}}
\end{gather}
where $dCov(\cdot)$ represents the distance covariance between two matrices; $dVar(\cdot)$ is the distance variance of each matrix.
For a more detailed calculation, refer to prior works~\cite{szekely2007measuring}.

\subsection{Model Optimization}
Having obtained the final representations for user $u$ and item $i$, we use inner product (\cf Equation~\eqref{equ:inner-product}) as the predictive function to estimate the likelihood of their interaction, $\hat{y}_{ui}$.
Thereafter, we use the pairwise BPR loss~\cite{BPRMF} to optimize the model parameters $\Theta=\{\Mat{u},\Mat{i}|u\in\Set{U},i\in\Set{I}\}$.
Specifically, it encourages the prediction of a user's historical items to be higher than that of unobserved items:
\begin{gather}\label{equ:bpr-loss}
    loss_{\text{BPR}}=\sum_{(u,i,j)\in\Set{O}}-\ln{\sigma(\hat{y}_{ui}-\hat{y}_{uj})} + \lambda\norm{\Theta}^{2}_{2},
\end{gather}
where $\Set{O}=\{(u,i,j)|(u,i)\in\Set{O}^{+},(u,j)\in\Set{O}^{-}\}$ denotes the training dataset involving the observed interactions $\Set{O}^{+}$ and unobserved counterparts $\Set{O}^{-}$; $\sigma(\cdot)$ is the sigmoid function; $\lambda$ is the coefficient controlling $L_{2}$ regularization.
During the training, we alternatively optimize the independence loss (\cf Equation~\eqref{equ:independence-loss}) and BPR loss (\cf Equation~\eqref{equ:bpr-loss}).

\subsection{Model Analysis}
In this subsection, we conduct model analysis, including the complexity analysis and DGCF's relations with existing models.


\subsubsection{\textbf{Model Size}}\label{sec:model-size}
While we slice the embeddings into $K$ chunks, the total embedding size remains the same as that set in MF and NGCF (\ie $d=64$).
That is, our DGCF involves no additional model parameters to learn, whole trainable parameters are $\{\Mat{u},\Mat{i}|u\in\Set{U},i\in\Set{I}\}$.
The model size of DGCF is identical to MF and lighter than NGCF that introduces additional transformation matrices.

\subsubsection{\textbf{Relation with LightGCN}}
LightGCN~\cite{lightgcn} can be viewed as a special case of DGCF with only-one intent representation and no independence modeling.
As the same datasets and experimental settings are used in these two models, we can directly compare DGCF with the empirical results reported in the LightGCN paper.
In terms of the recommendation accuracy, DGCF and LightGCN are in the same level.
However, benefiting from the disentangled representations, our DGCF has better interpretability since it can disentangle the latent intents in user representations (evidence from Table~\ref{tab:independence-impact}), and further exhibit the semantics of user intents (evidence from Section~\ref{sec:interpretability-of-representations}), while LightGCN fails to offer explainable representations.

\subsubsection{\textbf{Relation with Capsule Network}}
From the perspective of capsule networks~\cite{CapsuleNetwork,DBLP:conf/icann/HintonKW11}, we can view the chunked representations as the capsules and the iterative update rule as the dynamic routing.
However, distinct from typical capsule networks that adopt the routing across different layers, DGCF further incorporate the routing with the embedding propagation of GNNs, such that not only can the information be passed across layers, but also be propagated within the neighbors within the same layer.
As a result, our DGCF is able to distill useful information, that are relevant to intents, from multi-hop neighbors.

\subsubsection{\textbf{Relation with Multi-head Attetion Network}}
Although the intent-aware graphs can been seen as channels used in multi-head attention mechanism~\cite{Attention,GAT}, they have different utilities.
Specifically, multi-head attention is mainly used to stabilize the learning of attention network~\cite{GAT} and encourage the consistence among different channels; whereas, DGCF aims to collect different signals and leverage the independence module (\cf Section~\ref{sec:independence-modeling}) to enforce independence of intent-aware graphs.
Moreover, there is no information interchange between attention networks, while DGCF allows the intent-aware graphs to influence each other.

\subsubsection{\textbf{Relation with DisenGCN}}
DisenGCN~\cite{DisenGCN} is proposed to disentangle latent factors for graph representations, where the neighbor routing is also coupled with GNN models.
Our DGCF distinguishes from DisenGCN from several aspects: 1) DisenGCN fails to model the independence between factors, easily leading to redundant representations, while DGCF applies the distance correlation to achieve independence; 2) the embedding propagation in DisenGCN mixes the information from the ego node self and neighbors together; whereas, DGCF can purely distill information from neighbors;
and 3) DGCF's neighbor routing (\cf Section \ref{sec:graph-disentangling}) is more effective than that of DisenGCN (\cf Section~\ref{sec:overall-comparison}).
\section{Experiments}
\begin{table}[t]
    \caption{Statistics of the datasets.}
    \vspace{-10px}
    \label{tab:data-statistic}
    \resizebox{0.42\textwidth}{!}{
    \begin{tabular}{l|r|r|r|r}
    \hline
    \multicolumn{1}{c|}{Dataset} & \multicolumn{1}{c|}{\#Users} & \multicolumn{1}{c|}{\#Items} & \multicolumn{1}{c|}{\#Interactions} & \multicolumn{1}{c}{Density} \\ \hline\hline
    Gowalla & $29,858$ & $40,981$ & $1,027,370$ & $0.00084$ \\ \hline
    Yelp2018$^{*}$ & $31,668$ & $38,048$ & $1,561,406$ & $0.00130$ \\ \hline
    Amazon-Book & $52,643$ & $91,599$ & $2,984,108$ & $0.00062$ \\ \hline
    \end{tabular}}
    \vspace{-15px}
\end{table}

To answer the following research questions, we conduct extensive experiments on three public datasets:
\begin{itemize}[leftmargin=*]
    \item\textbf{RQ1:} Compared with present models, how does DGCF perform?
    \item\textbf{RQ2:} How different components (\eg layer number, intent number, independence modeling) affect the results of DGCF?
    \item\textbf{RQ3:} Can DGCF provide in-depth analyses of disentangled representations \wrt disentanglement of latent user intents and interpretability of representations?
\end{itemize}

\subsection{Experimental Settings}
\subsubsection{\textbf{Dataset Description}}
We use three publicly available datasets: Gowalla, Yelp2018$^{*}$, and Amazon-Book, released by NGCF~\cite{NGCF}.
Note that we revise Yelp2018 dataset, as well as the updated results of baselines, instead of the original reported in the NGCF paper~\cite{NGCF}\footnote{In the previous version of yelp2018, we did not filter out cold-start items in the testing set, and hence we rerun all methods.}.
We denote the revised as Yelp2018$^{*}$.
The statistics of datasets are summarized in Table~\ref{tab:data-statistic}.
We closely follow NGCF and use the same data split as NGCF.
In the training phase, each observed user-item interaction is treated as a positive instance, while we use negative sampling to randomly sample an unobserved item and pair it with the user as a negative instance.

\subsubsection{\textbf{Baselines}}
We compare DGCF with the state-of-the-art methods, covering the CF-based (MF), GNN-based (GC-MC and NGCF), disentangled GNN-based (DisenGCN), and disentangled VAE-based (MacridVAE):
\begin{itemize}[leftmargin=*]
    \item \textbf{MF}~\cite{BPRMF}: Such model treats user-item interactions as isolated data instances, and only uses ID embeddings as representations of users and items.
    
    \item \textbf{GC-MC}~\cite{GC-MC}: The method organizes historical user behaviors as a holistic interaction graph, and employs one GCN~\cite{GCN} encoder to generate representations. Note that only one-hop neighbors are involved.
    
    \item \textbf{NGCF}~\cite{NGCF}: This adopts three GNN layers on the user-item interaction graph, aiming to refine user and item representations via at most three-hop neighbors' information.

    \item \textbf{DisenGCN}~\cite{DisenGCN}: This is a state-of-the-art disentangled GNN model, which exploits the neighbor routing and embedding propagation to disentangle latent factors behind graph edges.
    

    \item \textbf{MacridVAE}~\cite{MacridVAE}: Such model is tailored to disentangle user intents behind user behaviors. In particular, it adopts $\beta$-VAE to estimate the generative process of a user's personal history, assuming that there are several latent factors affecting user behaviors, to achieved disentangled user representations. 
\end{itemize}

\subsubsection{\textbf{Evaluation Metrics.}}
To evaluate top-$N$ recommendation, we use the same protocols as NGCF~\cite{NGCF}:
recall@$N$ and ndcg@$N$\footnote{The previous implementation of ndcg metric in NGCF is slightly different from the standard definition, although reflecting the similar trends. We re-implement the ndcg metric and rerun all methods}, where $N$ is set as $20$ by default.
In the inference phase, we view historical items of a user in the test set as the positive, and evaluate how well these items are ranked higher than all unobserved ones.
The average results \wrt the metrics over all users are reported.

\subsubsection{\textbf{Parameter Settings.}}
We implement the DGCF model in Tensorflow.
For a fair comparison, we tune the parameter settings of each model.
In particular, we directly copy the best performance of MF, GC-MC, and NGCF reported in the original paper~\cite{NGCF}.
As for DisenGCN and DGCF, we fix the embedding size $d$ as $64$ (which is identical to MF, GC-MC, and NGCF), use Adam~\cite{Adam} as the optimizer, initialize model parameters with Xarvier~\cite{Xarvier}, and fix the iteration number $T$ as $2$.
Moreover, a grid search is conducted to confirm the optimal settings --- that is, the learning rate is searched in $\{0.001,0.005,0.0005,0.0001\}$, and the coefficients $\lambda$ of $L_{2}$ regularization term is tuned in $\{10^{-3},10^{-4},10^{-5}\}$.
For MacridVAE~\cite{MacridVAE}, we search the number of latent factors being disentangled in $\{2,4,8,16\}$, and tune the factor size in $\{50,55,60,65,70\}$.

Without specification, unique hyperparameters of DGCF are set as: $L=1$ and $K=4$.
We study the number of graph disentangling layer $L$ in $\{0,1,2,3\}$ and the number of latent intents $K$ in $\{2,4,8,16\}$, and report their influences in Sections~\ref{sec:layer-number-impact} and~\ref{sec:intent-number-impact}, respectively.

\begin{table}[t]
    \caption{Overall Performance Comparison.}
    \vspace{-10px}
    \label{tab:overall-performance}
    \resizebox{0.47\textwidth}{!}{
    \begin{tabular}{l|cc|cc|cc}
    \hline
     & \multicolumn{2}{c|}{Gowalla} & \multicolumn{2}{c|}{Yelp2018$^{*}$} & \multicolumn{2}{c}{Amazon-Book} \\
     & recall & ndcg & recall & ndcg & recall & ndcg \\ \hline\hline
    MF & 0.1291 & 0.1109 & 0.0433 & 0.0354 & 0.0250 & 0.0196 \\
    GC-MC & 0.1395 & 0.1204 & 0.0462 & 0.0379 & 0.0288 & 0.0224 \\ 
    NGCF & 0.1569 & \underline{0.1327} & 0.0579 & 0.0477 & 0.0337 & 0.0266 \\ \hline
    DisenGCN & 0.1356 & 0.1174 & 0.0558 & 0.0454 & 0.0329 & 0.0254 \\
    MacridVAE & \underline{0.1618} & 0.1202 & \underline{0.0612}  & \underline{0.0495} & \underline{0.0383} & \underline{0.0295} \\ \hline
    DGCF-1 & $\Mat{0.1794}^{*}$ & $\Mat{0.1521}^{*}$ & $\Mat{0.0640}^{*}$ & $\Mat{0.0522}^{*}$ & $\Mat{0.0399}^{*}$ & $\Mat{0.0308}^{*}$ \\ \hline\hline
    \%improv. & 10.88\% & 14.62\% & 4.58\% & 5.45\% & 4.17\% & 4.41\% \\
    $p$-value & 6.63e-8 & 3.10e-7 & 1.75e-8 & 4.45e-9 & 8.26e-5 & 7.15e-5 \\ \hline
    \end{tabular}}
    \vspace{-15px}
\end{table}

\subsection{Performance Comparison (RQ1)}\label{sec:overall-comparison}
We report the empirical results of all methods in Table~\ref{tab:overall-performance}.
The improvements and statistical significance test are performed between DGCF-1 with the strongest baselines (highlighted with underline). Analyzing such performance comparison, we have the following observations:
\begin{itemize}[leftmargin=*]
    \item Our proposed DGCF achieves significant improvements over all baselines across three datasets. In particular, its relative improvements over the strongest baselines \wrt recall@$20$ are $10.88\%$, $4.58\%$, and $4.17\%$ in Gowalla, Yelp2018$^{*}$, and Amazon-Book, respectively. This demonstrates the high effectiveness of DGCF. We attribute such improvements to the following aspects --- 1) by exploiting diverse user-item relationships, DGCF is able to better characterize user preferences, than prior GNN-based models that treat the relationships as uniform edges; 2) the disentangling module models the representations at a more granular level of user intents, endowing the recommender better expressiveness; and 3) embedding propagation mechanism can more effectively distill helpful information from one-hop neighbors, than one-layer GC-MC and three-layer NGCF.
    
    \item Jointly analyzing the results across three datasets, we find that the improvements on Amazon-Book is much less than that on the others. This might suggest that, purchasing books is a simper scenario than visiting location-based business. Hence, the user intents on purchasing books are less diverse.
    
    \item MF performs poor on three datasets. This indicates that, modeling user-item interactions as isolated data instance could ignore underlying relationships among users and items, and easily lead to unsatisfactory representations.
    
    \item Compared with MF, GC-MC and NGCF consistently achieve better performance on three datasets, verifying the importance of user-item relationships. They take the one-hop neighbors (\ie reflecting behavioral similarity of users) and third-hop neighbors (\ie carrying collaborative signals) into representations.
    
    \item DisenGCN substantially outperforms GC-MC in most cases. It is reasonable since user intents are explicitly modeled as factors being disentangled in DisenGCN, which offer better guide to distill information from one-hop neighbors. However, its results are worse than that of DGCF. Possible reasons are that 1) its routing mechanism only uses node affinity to adjust the graph structure, without any priors to guide; and 2) many operations (\eg nonlinear transformation) are heavy to CF~\cite{SGC}. This suggests its suboptimal disentanglement.

 
    \item MacridVAE serves as the strongest baseline in most cases. This justifies the effectiveness of estimating personal history via a generative process, and highlights the importance of the disentangled representations.
\end{itemize}

\subsection{Study of DGCF (RQ2)}
Ablation studies on DGCF are also conducted to investigate the rationality and effectiveness of some designs --- to be more specific, how the number of graph disentangling layers, the number of latent user intents, and independence modeling influence the model. 

\begin{table}[t]
    \caption{Impact of Layer Number ($L$).}
    \vspace{-10px}
    \label{tab:layer-number-impact}
    \resizebox{0.47\textwidth}{!}{
    \begin{tabular}{l|cc|cc|cc}
    \hline
     & \multicolumn{2}{c|}{Gowalla} & \multicolumn{2}{c|}{Yelp2018$^{*}$} & \multicolumn{2}{c}{Amazon-Book} \\
     & recall & ndcg & recall & ndcg & recall & ndcg \\ \hline\hline
    DGCF-1 & 0.1794 & 0.1521 & 0.0640 & 0.0522 & 0.0399 & 0.0308 \\ 
    DGCF-2 & 0.1834 & 0.1560 & 0.0653 & 0.0532 & 0.0422 & $\Mat{0.0324}^{*}$ \\ 
    DGCF-3 & $\Mat{0.1842}^{*}$ & $\Mat{0.1561}^{*}$ & $\Mat{0.0654}^{*}$ & $\Mat{0.0534}^{*}$ & $\Mat{0.0422}^{*}$ & 0.0322  \\ \hline
    \end{tabular}}
    \vspace{-15px}
\end{table}

\begin{figure}[t]
    \centering
    \subfigure[Intent Impact in Gowalla]{
    \label{fig:intent-impact-gowalla}\includegraphics[width=0.227\textwidth]{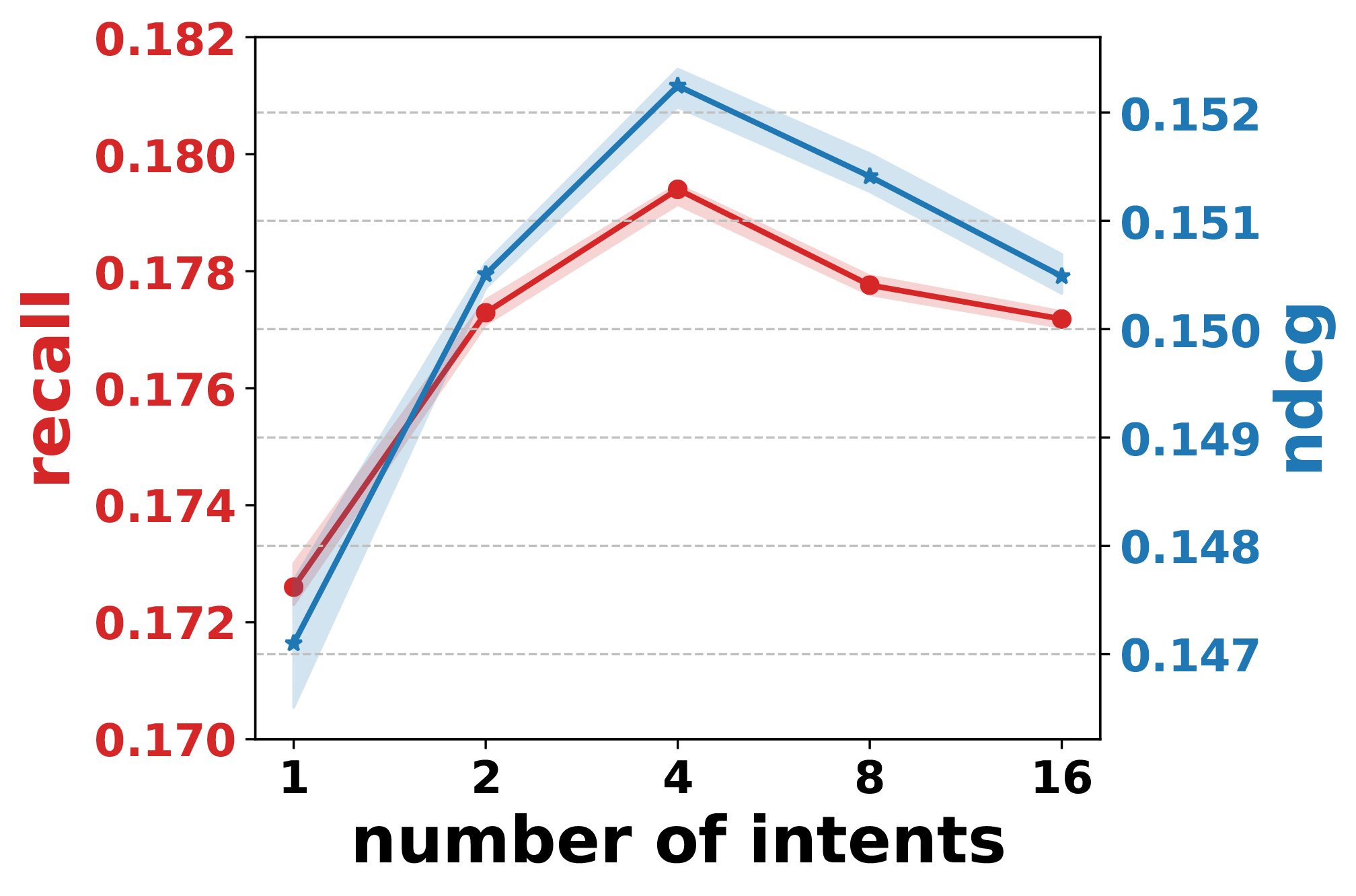}}
    \subfigure[Intent Impact in Yelp208$^{*}$]{
    \label{fig:intent-impact-yelp2018}\includegraphics[width=0.227\textwidth]{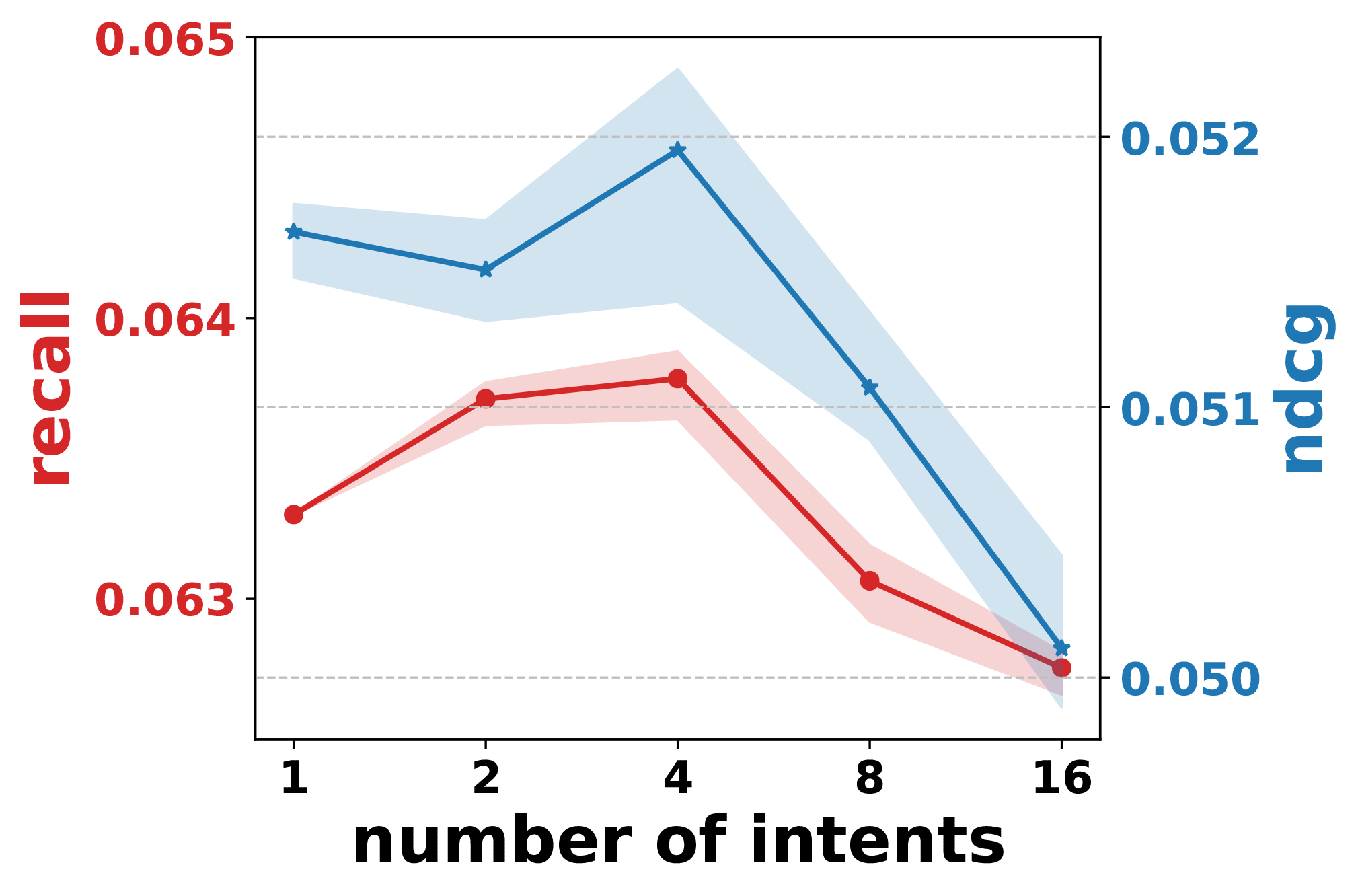}}
    \vspace{-15pt}
    \caption{Impact of Intent Number ($K$). Best Viewed in Color.}
    \label{fig:intent-impact}\vspace{-15pt}
\end{figure}

\subsubsection{\textbf{Impact of Model Depth.}}\label{sec:layer-number-impact}
The graph disentangling layer is at the core of DGCF, which not only disentangles user intents, but also collects information pertinent to individual intents into intent-aware representations. Furthermore, stacking more layers is able to distill the information from multi-hop neighbors. Here we investigate how the number of such layers, $L$, affects the performance.
Towards that, we search $L$ in the range of $\{1,2,3\}$ and summarize the empirical results in Table~\ref{tab:layer-number-impact}. Here we use DGCF-1 to denote the recommender with one disentangling layer, and similar notations for others.
We have several findings:
\begin{itemize}[leftmargin=*]
    \item Clearly, increasing the model depth is capable of endowing the recommender with better representation ability. In particular, DGCF-2 outperforms DGCF-1 by a large margin. This makes sense since DGCF-1 just gathers the signals from one-hop neighbors, while DGCF-2 considers multi-hop neighbors.
    
    \item While continuing stacking more layer beyond DGCF-2 gradually improves the performance of DGCF-3, the improvements are not that obvious. This suggests that the second-order connectivity might be sufficient to distill intent-relevant information. Such observation is consistent to NGCF~\cite{NGCF}.
    
    \item We compare the results across Tables~\ref{tab:overall-performance} and~\ref{tab:layer-number-impact} and find that DGCF with varying depths is consistently superior to other baselines. This again verifies the effectiveness of embedding propagation with neighbor routing.
\end{itemize}

\subsubsection{\textbf{Impact of Intent Number.}}\label{sec:intent-number-impact}
To study the influence of intent number, we vary $K$ in range of $\{1,2,4,8,16\}$ and demonstrate the performance comparison on Gowalla and Yelp$^{*}$ datasets in Figure~\ref{fig:intent-impact}.
There are several observations:
\begin{itemize}[leftmargin=*]
    \item Increasing the intent number from $1$ to $4$ significantly enhances the performance. In particular, DGCF performs the worst when $K=1$, indicating that only uniform relationship is not sufficient to profile behavioral patterns of users. This again justifies the rationality of disentangling user intents.
    
    \item Interestingly, the recommendation performance drops when then intent number reaches $8$ and $16$. This suggests that, although benefiting from the proper intent chunks, the disentanglement suffers from too fine-grained intents. One possible reason is that the chunked representations only with the embedding size of $4$ (\eg $\frac{d}{K}=4$ when $K=16$) have limited expressiveness, hardly describing a complete pattern behind user behaviors.
 
    \item Separately analyzing the performance in Gowalla and Yelp208$^{*}$, we find that the influences of intent number are varying. We attribute this to the differences between scenarios. To be more specific, Gowalla is a location-based social networking service which provides more information, such as social marketing, as the intents than Yelp2018$^{*}$.
    This indicates that user behaviors are driven by different intents for difference scenarios.
\end{itemize}


\begin{table}[t]
    \caption{Distance correlation of different methods (the lower the better).}
    \vspace{-10px}
    \label{tab:independence-impact}
    \resizebox{0.31\textwidth}{!}{
    \begin{tabular}{l|ccc}
    \hline
     & Gowalla & Yelp$^{*}$ & Amazon-Book \\ \hline\hline
    MF & 0.2332 & 0.2701 & 0.2364 \\ 
    NGCF & 0.2151 & 0.2560 & 0.1170 \\ \hline
    DGCF-1 & 0.1389 & 0.1713 & 0.1039 \\ 
    DGCF-2 & $\Mat{0.0920}^{*}$ & $\Mat{0.1217}^{*}$ & $\Mat{0.0751}^{*}$ \\ \hline
    \end{tabular}}
    \vspace{-12px}
\end{table}


\begin{table}[]
    \caption{Impact of Independence \wrt recall in Yelp2018$^{*}$.}
    \vspace{-10px}
    \label{tab:independence-impact2}
    \resizebox{0.28\textwidth}{!}{
    \begin{tabular}{l|c|c|c}
    \hline
     & DGCF-1 & DGCF-2 & DGCF-3 \\ \hline\hline
    w/o ind & 0.0637 & 0.0649 & 0.0650 \\ \hline
    w/ ind & 0.0640 & 0.0653 & 0.0654 \\ \hline
    \end{tabular}}
    \vspace{-15px}
\end{table}

\begin{figure*}[t]
    \centering
	\includegraphics[width=0.97\textwidth]{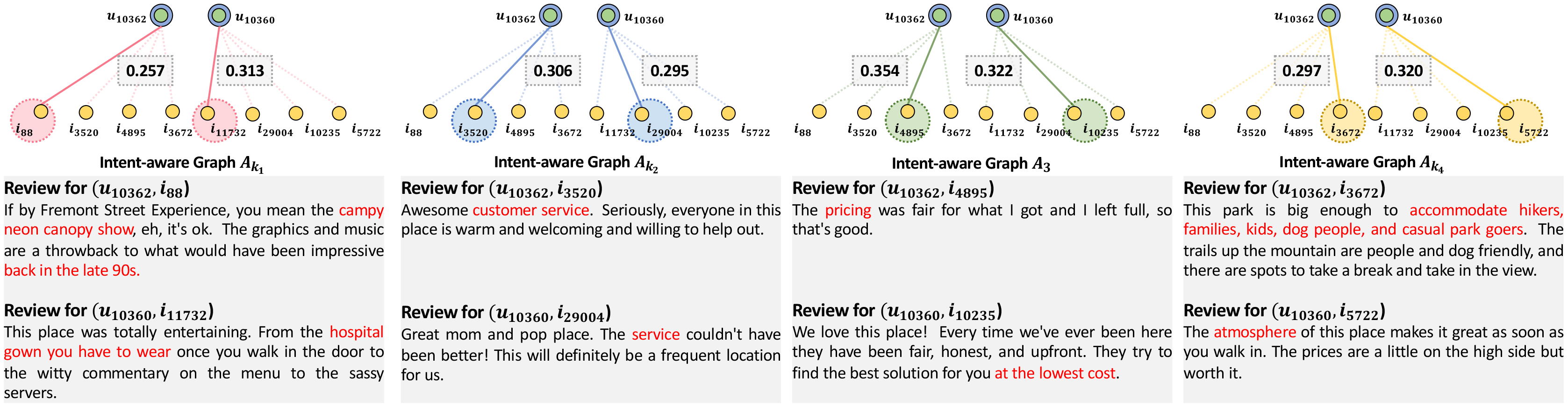}
	\vspace{-10pt}
	\caption{Illustration of the proposed disentangled graph collaborative filtering framework. Best viewed in color.}
	\label{fig:case-study}
	\vspace{-15pt}
\end{figure*}

\begin{figure}[t]
    \centering
    \subfigure[Disentanglement \wrt recall.]{
    \label{fig:disentanglement-recall}\includegraphics[width=0.23\textwidth]{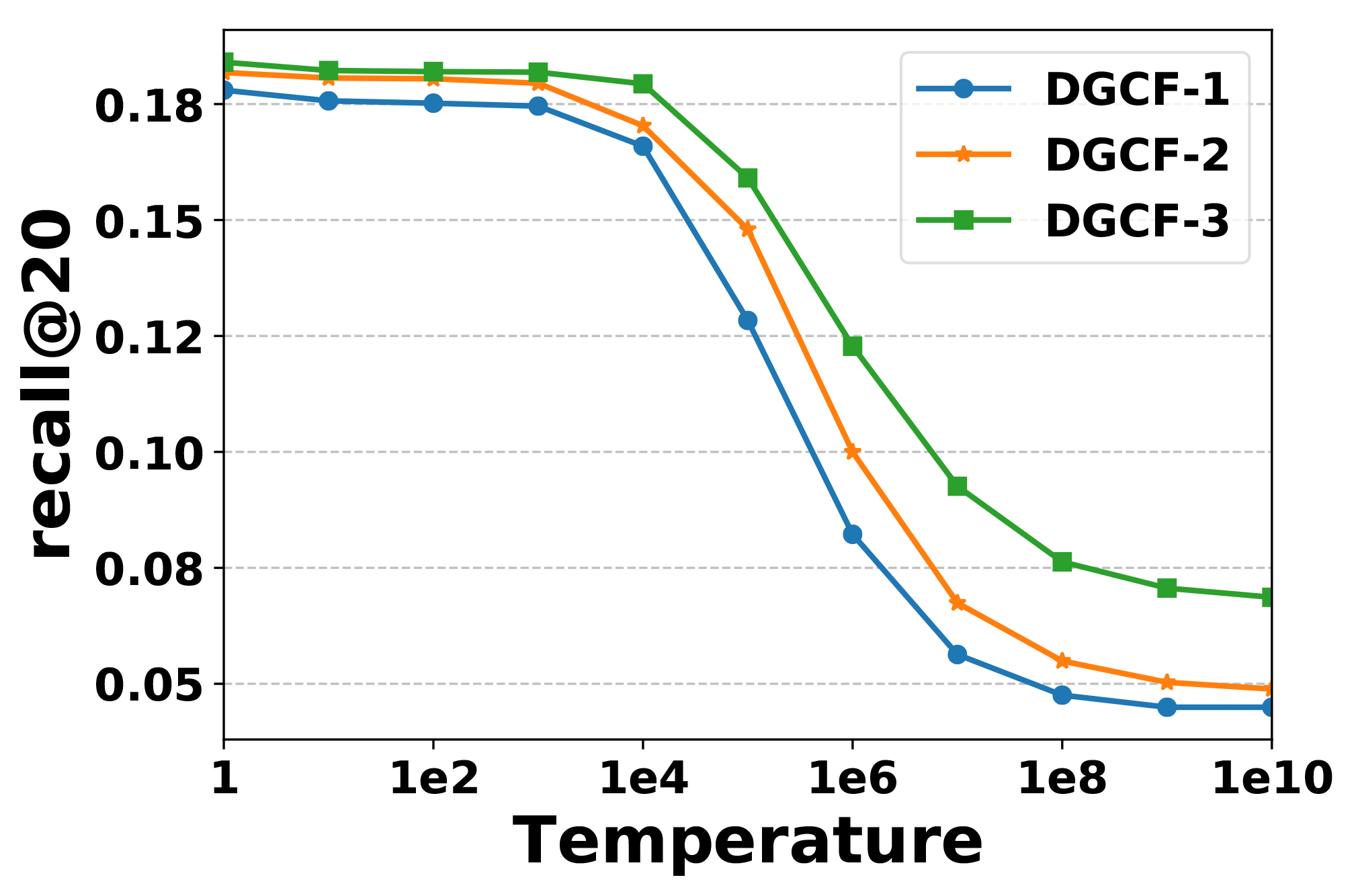}}
    \subfigure[Disentanglement \wrt ndcg.]{
    \label{fig:disentanglement-ndcg}\includegraphics[width=0.23\textwidth]{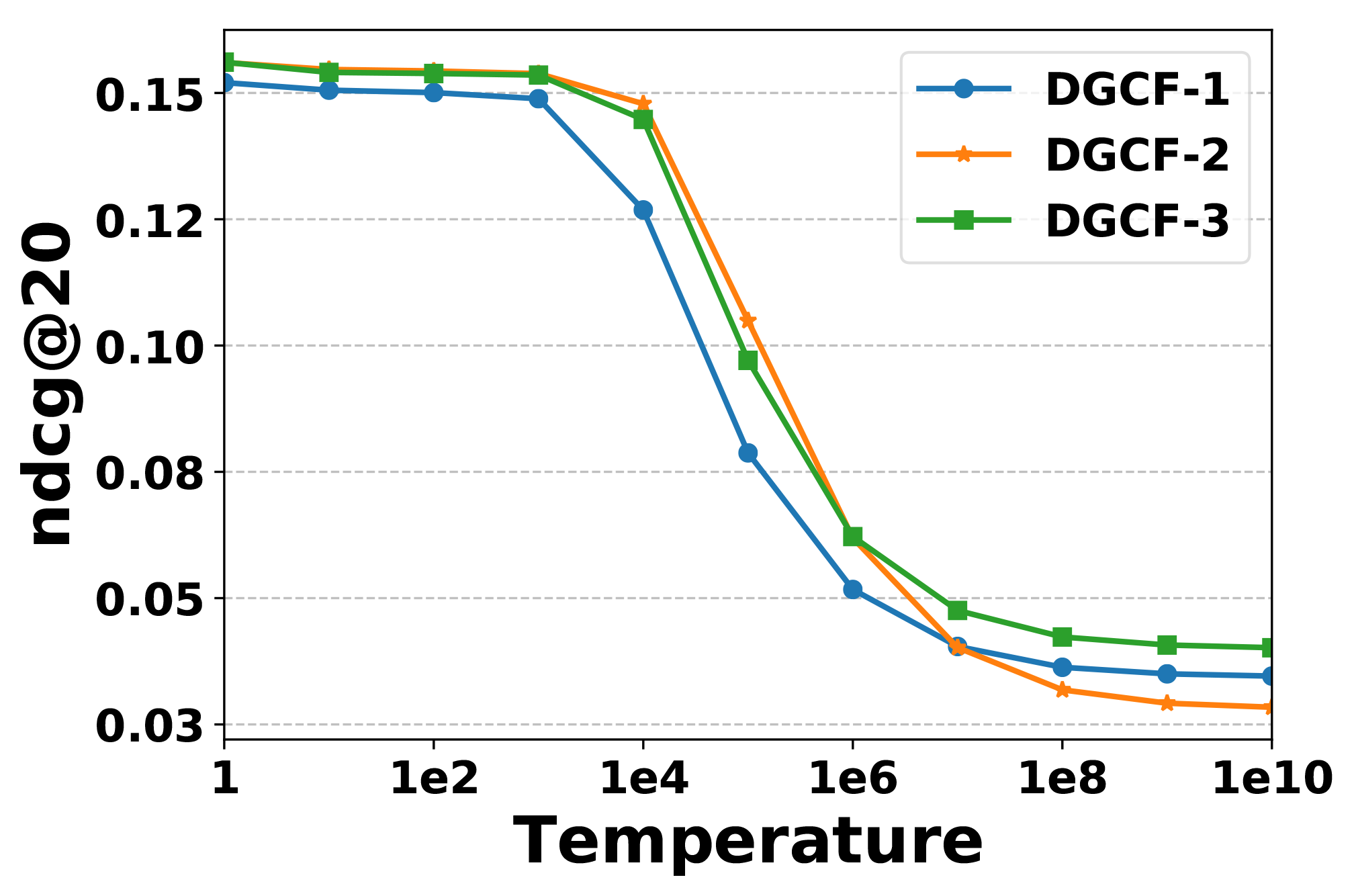}}
    \vspace{-15pt}
    \caption{Disentanglement of Intents. Best Viewed in Color.}
    \label{fig:disentanglement}\vspace{-15pt}
\end{figure}

\subsubsection{\textbf{Impact of Independence Modeling.}}\label{sec:independence-impact}
As introduced in Equation~\eqref{equ:independence-loss} distance correlation is a statistic measure to quantify the level of independence, we hence select MF, NGCF, and DGCF to separately represent three fashions to model user-item relationships --- isolated data instance, holistic interaction graph, and intent-aware interaction graphs, respectively.
We report the result comparison \wrt distance correlation in Table~\ref{tab:independence-impact}.
Furthermore, conduct another ablation study to verify the influence of independence modeling.
To be more specific, we disable this module in DGCF-1, DGCF-2, and DGCF-3 to separately build the variants DGCF-1$_{\text{ind}}$, DGCF-2$_{\text{ind}}$, and DGCF-3$_{\text{ind}}$.
We show the results in Table~\ref{tab:independence-impact2}. There are some observations:
\begin{itemize}[leftmargin=*]
    \item A better performance is substantially coupled with a higher level of independence across the board. In particular, focusing on one model group, the intents of DGCF$_{\text{ind}}$ are more highly entangled than that of DGCF; meanwhile, DGCF is superior to its variants \wrt recall in Gowalla and Yelp2018 datasets. This reflects the correlation between the performance and disentanglement, which is consistent to the observation in~\cite{MacridVAE}.
    
    \item When analyzing the comparison across model groups, we find that the performance drop caused by removing the independence modeling from DGCF-2 is more obvious in DGCF-1. This suggests that the independence regularizer is more beneficial to the deeper models, leading to better intent-aware representations.
\end{itemize}

\subsection{In-depth Analysis (RQ3)}
In what follows, we conduct experiments to get deep insights into the disentangled representations \wrt the disentanglement of intents and interpretability of representations.

\subsubsection{\textbf{Disentanglement of Intents.}}\label{sec:disentanglement-of-intents}
Disentangled representation learning aims to generate factorial representations, where a change in one factor is relatively invariant to changes in other factors~\cite{RepresentationLearning}.
We hence investigate whether the intent-aware representations of DGCF match this requirement well, and get deep insights into the disentanglement.
Towards this end, we introduce a temperature factor, $\tau$, into the graph disentangling module.
Specifically, for each user/item factorial representation, we first select a particular intent $k$, whose $A_{k}(u,i)$ is the smallest in $K$ intents, and then use $\tau$ to reassign it with a smaller weight $\frac{A_{k}(u,i)}{\tau}$, while remaining the other intents unchanged.
Wherein, $\tau$ is searched in the range of $\{10^{0},10^{1},10^{2},\cdots,10^{10}\}$.
Through this way, we study its influence on recommendation performance.
The results in Gowalla dataset is shown in Figure~\ref{fig:disentanglement}.
We have several interesting observations:
\begin{itemize}[leftmargin=*]
    \item The discovered intents serve as \emph{vitamin} to the disentangled representations. That is, even the intents in small quantities (\ie when $\tau$ ranges from $10^{0}$ to $10^{3}$) are beneficial to the representation learning. This also verifies the independence of intents: the changes on the specific intent almost have no influence on the others, still leading to comparable performances to the optimal one.
    
    \item The \emph{vitamin} also means that the deficiency of any intent hinders the results severely. When $\tau$ is larger than $10^{4}$, the performance drops quickly, which indicates that lack of intents (whose influences is close to zero) negatively affects the model. Moreover, this verifies that one intent cannot be inferred by the others.

\end{itemize}

\subsubsection{\textbf{Interpretability of Representations.}}\label{sec:interpretability-of-representations}
We now explore the semantics of the explored intents.
Towards this end, in Yelp2018$^{*}$ datasets, we use side information (\ie user reviews) which possibly contain some cues for why users visit the local business (\ie self-generated explanations on their behaviors).
In particular, we randomly selected two users $u_{10362}$ and $u_{10360}$, and disentangled the explored intents behind each historical user-item interactions, based on the learned intent-aware graphs.
Thereafter, for each interaction, we coupled its review with the intent with the highest confidence score.
For example, conditioned on the intent $k_{1}$, interaction $(u_{10360},i_{11732})$ has the hightest score $0.313$ among $u_{10362}$'s history.
Figure~\ref{fig:case-study} shows the post-hoc explanations of intents, and we have the following findings:
\begin{itemize}[leftmargin=*]
    \item Jointly analyzing reviews for the same intent, we find that, while showing different fine-grained preferences, they are consistent with some high-level concepts. For example, intent $k_{1}$ contributes the most to interactions $(u_{10362},i_{88})$ and $(u_{10360},i_{11732})$, which suggests its high confidence of being the intents behind these behaviors. We hence analyze the reviews of interactions, and find that they are about users' special interests: \emph{the late 90s} for $u_{10362}$, and \emph{hospital gown} for $u_{10360}$.
    
    \item DGCF is able to discover user intents inherent in historical behaviors. Examining the corresponding reviews, we summarize the semantics (high-level concept) of these four intents $k_{1}$, $k_{2}$, $k_{3}$, and $k_{4}$ as \emph{interesting matching}, \emph{service}, \emph{price\&promotion}, and \emph{passing the time}, respectively. This verifies the rationality of our assumptions: user behaviors are driven by multiple intents with varying contributions.

    \item This inspires us to go beyond user-item interactions and consider extra information to model user intents in an explicit fashion. We plan to incorporate psychology knowledge in future work.
\end{itemize}

\section{Conclusion And Future Work}
In this work, we exhibited user-item relationships at the granularity of user intents and disentangled these intents in the representations of users and items.
We devised a new framework, DGCF, which utilizes the graph disentangling module to iteratively refine the intent-aware interaction graphs and factorial representations.
We further introduce the independence modeling module to encourage the disentanglement.
We offer deep insights into the DGCF \wrt effectiveness of recommendation, disentanglement of user intents, and interpretability of factorial representations.

Learning disentangled user intents is an effective solution to exploit diverse relationships among users and items, and also helps greatly to interpret their representations.
This work shows initial attempts towards interpretable recommender models.
In future work, we will involve side information, such as conversation history with users~\cite{DBLP:conf/wsdm/Lei0MWHKC20}, item knowledge~\cite{KPRN,KGAT,KGPolicy,liumeng,DBLP:journals/tip/HongLCTWT17}, and user reviews~\cite{DBLP:conf/sigir/LiQPQDW19,DBLP:conf/www/ChengDZK18}, or conduct psychology experiments~\cite{lu2019quality} to establish the ground truth on user intents and further interpret disentangled representations better.
Furthermore, we would like to explore the privacy and robustness of the factorial representations, with the target of avoiding the leakage of sensitive information.

\begin{acks}
    This research is part of NExT++ research, which is supported by the National Research Foundation, Singapore under its International Research Centres in Singapore Funding Initiative. It is also supported by the National Natural Science Foundation of China (U19A2079).
\end{acks}


\bibliographystyle{ACM-Reference-Format}
\balance
\bibliography{ms}
\balance

\end{document}